\documentclass[sigconf]{acmart} 
\AtBeginDocument{%
  }

\setcopyright{acmlicensed}
\copyrightyear{2025}
\acmYear{2025}
\acmDOI{XXXXXXX.XXXXXXX}
\acmConference[ICNS3 '25]{International Conference on ns-3}{August 19--21, 2025}{Osaka, Japan}

\usepackage{amsmath, amsfonts}
\usepackage{graphicx}
\usepackage{float}
\usepackage{hyperref}
\usepackage{cleveref}
\usepackage{caption}
\usepackage{siunitx}
\usepackage{pbox}
\usepackage{subcaption}

\usepackage[absolute,overlay]{textpos}
\usepackage[most]{tcolorbox}

\usepackage{tikz}
\usetikzlibrary{positioning, arrows.meta, calc, fit, matrix, patterns}

\graphicspath{{./images}}
\begin{document}
\title{Design and Evaluation of IEEE 802.11ax Uplink Orthogonal Frequency Division Multiple Random Access in ns-3}


\author{Douglas Dziedzorm Agbeve}
\affiliation{%
  \institution{IDLab, University of Antwerp -- IMEC}
  \city{Antwerp}
  \country{Belgium}}
\email{douglas.agbeve@uantwerpen.be}

\author{Andrey Belogaev}
\affiliation{%
  \institution{IDLab, University of Antwerp -- IMEC}
  \city{Antwerp}
  \country{Belgium}}
\email{andrei.belogaev@uantwerpen.be}

\author{Jeroen Famaey}
\affiliation{%
 \institution{IDLab, University of Antwerp -- IMEC}
 \city{Antwerp}
\country{Belgium}}
\email{jeroen.famaey@uantwerpen.be}

\renewcommand{\shortauthors}{D.D. Agbeve et al.}

\begin{abstract}
  Wi-Fi networks have long relied on the Enhanced Distributed Channel Access
  (EDCA) mechanism, allowing stations to compete for transmission opportunities.
  However, as networks become denser and emerging applications demand lower
  latency and higher reliability, the limitations of EDCA — such as overhead due
  to contention and collisions — have become more pronounced. To address these
  challenges, Orthogonal Frequency Division Multiple Access (OFDMA) has been
  introduced in Wi-Fi, enabling more efficient channel utilization through
  scheduled resource allocation. Furthermore, Wi-Fi 6 defines Uplink Orthogonal
  Frequency Division Multiple Random Access (UORA), a hybrid mechanism that
  combines both scheduled and random access, balancing efficiency and
  responsiveness in resource allocation. Despite significant research on UORA,
  most studies rely on custom simulators that are not publicly available,
  limiting reproducibility and preventing validation of the presented results.
  The only known open-source UORA implementation in the ns-3 simulator exhibits
  key limitations, such as usage of the same trigger frame (TF) to schedule
  resources for buffer status reports and data transmissions, and lack of
  signaling for UORA configuration. In this paper, we present a fully
  standard-compliant and open source UORA implementation that is compatible with
  ns-3 version 3.38, addressing these limitations to improve resource
  allocation efficiency and adaptability. This implementation enables more
  accurate and flexible evaluation of UORA, fostering future research on Wi-Fi
  resource allocation strategies.
\end{abstract}




\maketitle
\begin{textblock*}{\textwidth}(2.5cm, 25.2cm) 
    \begin{tcolorbox}[colframe=black,
        colback=white,
        boxrule=0.4pt,
        width=.9\textwidth,
        boxsep=0pt,
        left=1pt, right=1pt, top=1pt, bottom=1pt]
        \normalsize \textbf{\copyright~Agbeve et al. | ACM 2025. This is the author's
            version of the work. It is posted here for your personal use. Not
            for redistribution. The definitive Version of Record was published
        in ---.}
    \end{tcolorbox}
\end{textblock*}
\section{Introduction \label{sec:intro}}

For a long time, the channel access in Wi-Fi networks has been dominated by a random
channel access mechanism called Enhanced Distributed Channel Access
(EDCA). This mechanism allows multiple Wi-Fi stations to compete for the channel,
with the possibility that multiple stations will transmit at the same time,
leading to a \emph{collision}, and consequent unsuccessful decoding of the data
packets. In such cases, Wi-Fi stations will use the EDCA procedure again to gain
access to the channel for retransmissions. With the evolution of Wi-Fi networks,
multiple factors came into play. Firstly, networks have become denser, which leads
to a higher number of devices competing for the channel and therefore more collisions. Secondly, new
applications have emerged, such as Extended Reality (XR) and Industrial
Automation, which require significantly lower delays with
higher guarantees. Finally, the new Wi-Fi standards have significantly increased
the amount of bandwidth, from 20 MHz in the early Wi-Fi standards to hundreds of
MHz in the latest ones. As a consequence, the IEEE 802.11ax amendment, also
known as Wi-Fi 6, introduced scheduled channel access called Orthogonal
Frequency Division Multiple Access (OFDMA) to Wi-Fi networks~\cite{80211ax}.
Unlike earlier Wi-Fi standards, where a station occupies the entire
channel bandwidth during transmission, OFDMA divides the available bandwidth
into multiple Resource Units (RUs), which are assigned to different stations.

Using OFDMA, the access point (AP) can assign RUs to stations (STAs) associated
with it. Wi-Fi supports both downlink (DL) and uplink (UL) OFDMA transmissions. The new channel
access mechanism opens up many opportunities for efficient centralized resource
management. Moreover, it significantly reduces the contention in uplink. Indeed,
with UL OFDMA the STAs can stop competing for the channel, because the AP can
now assign resources to them. This way, a significant part of the overhead
related to back-off timers and retransmissions after collisions can be
eliminated. To assign resources in uplink, the AP should be aware of the status
of the buffers on the STAs. For that, the AP can also assign RUs for polling,
i.e., for buffer status reports (BSRs) transmitted from STAs in uplink. With
rising number of the STAs it becomes increasingly more complex for the AP to
schedule RUs on time to avoid ineffective use of RUs, i.e., assignment of RUs to
STAs with empty buffers while STAs with packets in their buffers are waiting for
too long~\cite{agbeve2025a2p}. To overcome this issue, Wi-Fi 6 defines a hybrid
approach called \emph{Uplink Orthogonal Frequency Division Multiple Random
Access (UORA)}, which shares the properties of both random and scheduled access
mechanisms. Specifically, the AP divides the RUs into two types: Scheduled
Access (SA) RUs and Random Access (RA) RUs. Then the AP transmits a special
frame called trigger frame (TF), which contains information about which RUs are
assigned to which STAs. Upon receiving the TF, the scheduled STAs transmit using
their assigned SA RUs in a contention-free manner. In contrast, all STAs can
compete for RA RUs through a contention-based mechanism, which is described in
detail in Section~\ref{sec:uora_descr}. Using UORA, the AP can maintain the
balance between efficient scheduling of resources to the STAs with known buffer
sizes and polling the STAs for information about new packets.

Multiple research papers~\cite{avdotin2019enabling, avdotin2020resource,
bhattarai2019uplink, chen2020scheduling, jin2024enhancing, kim2021ofdma,
kosek2022efficient, kosek2022improving, rehman2023collision, xie2020multi} have
studied UORA from different perspectives. The majority of the papers propose
various extensions to UORA for improvement in terms of resource utilization,
number of collisions, and energy efficiency. Most of these papers rely on
simulation for the performance evaluation of the proposed algorithms. However,
the simulation results have been obtained in custom simulators, which are not
publicly available for testing and validation. To the best of our knowledge, the
only implementation of UORA in a major trustworthy open source network simulator
was developed by Naik \textit{et al.}~\cite{1_Naik}. The implementation makes
use of the ns-3 simulator~\cite{riley2010ns}, and the results have been
validated against a mathematical model. However, this implementation has several
shortcomings. Firstly, due to its design, only one trigger frame (TF) is used in
UL OFDMA transmission, meaning the same resources are allocated for buffer
status reporting and data transmission, resulting in inefficient resource
utilization. Secondly, the signaling required to configure the UORA parameter
set on STAs has not been implemented, thereby restricting on-the-fly parameter
configuration adjustments.
In this paper, we present a fully standard-compliant implementation
of UORA compatible with ns-3 version 3.38, which incorporates a standalone
scheduler designed to maximize the full potential of UORA by decoupling resource
allocation for buffer status reports and data transmissions.

The remainder of the paper is organized as follows. In Section~\ref{sec:uora_descr},
we provide a detailed description of the UORA operation. We also give a brief overview of the state of the
art of research related to UORA performance evaluation and optimization. In
Section~\ref{sec:uora_impl}, we explain the key components of the proposed
implementation in the ns-3 simulator. In Section~\ref{sec:validation}, we show the
results of our implementation's validation. Finally, in
Section~\ref{sec:conclusion}, we conclude the paper.

\section{Background}
\label{sec:uora_descr}

\tikzstyle{sta_freq} = [rectangle, draw, align=flush center, thin,
                        minimum height = .5cm, minimum width = 2.6cm, outer sep =
                        0pt]
\tikzstyle{ap_freq} = [rectangle, draw, align=flush center, thin,
                        minimum height = .5cm, minimum width = 1cm]
\tikzstyle{line} = [-{latex[length=7mm, width=5mm]}, thin,
                        draw,color=black!100]
\tikzstyle{sifs} = [{latex[length=7mm, width=5mm]}-{latex[length=7mm,
width=5mm]}, thin, draw,color=black!100]

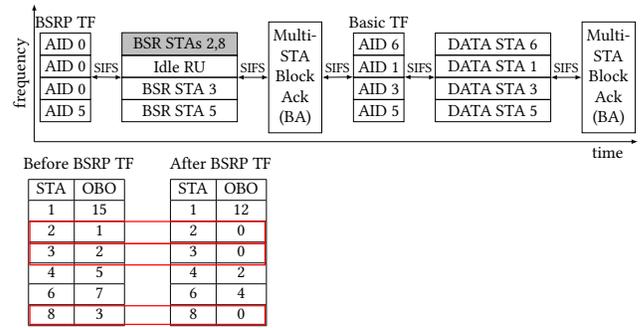
\begin{figure}[tb]
    \Description{Empty}
    \centering
        \scalebox{0.59}{
            \begin{tikzpicture}[node distance = .7cm, on
                grid, auto, label position=center, align = flush center]
                \node[ap_freq, xshift = -1.3cm, yshift = -.3cm, font=\Large](BSRP_TF1) {AID 0};
                \node at (BSRP_TF1.north) [above, font=\Large ] {BSRP TF};
                \node[ap_freq, below = of BSRP_TF1, yshift = .2cm, font=\Large](BSRP_TF2)
                    {AID 0};
                \node[ap_freq, below = of BSRP_TF2, yshift = .2cm, font=\Large](BSRP_TF3)
                    {AID 0};
                \node[ap_freq, below = of BSRP_TF3, yshift = .2cm, font=\Large](BSRP_TF4)
                    {AID 5};
                \node[sta_freq, right = of BSRP_TF1.east, xshift =0cm,
                    fill=gray!50, font=\Large](BSR_STA1){BSR STAs 2,8};
                \node[sta_freq, below = of BSR_STA1, yshift =
                    .2cm, font=\Large](BSR_STA2){Idle RU};
                \node[sta_freq, below = of BSR_STA2, yshift = .2cm, font=\Large](BSR_STA3){BSR STA 3};
                \node[sta_freq, below = of BSR_STA3, yshift = .2cm, font=\Large](BSR_STA4){BSR STA 5};
                \node[ap_freq, right = of BSR_STA1.east, minimum height = 2.5cm,
                    yshift = -.75cm, font=\Large](BlockACK1)
                    {Multi-\\STA\\Block\\Ack\\(BA)};
                \node[ap_freq, right = of BlockACK1.east, yshift =
                    .75cm, font=\Large](BASIC_TF1) {AID 6};
                \node at (BASIC_TF1.north) [above, font=\Large] {Basic TF};
                \node[ap_freq, below = of BASIC_TF1, yshift = .2cm, font=\Large](BASIC_TF2)
                    {AID 1};
                \node[ap_freq, below = of BASIC_TF2, yshift = .2cm, font=\Large](BASIC_TF3)
                    {AID 3};
                \node[ap_freq, below = of BASIC_TF3, yshift = .2cm, font=\Large](BASIC_TF4)
                    {AID 5};
                 \node[sta_freq, right = of BASIC_TF1.east, font=\Large](DATA_STA1){DATA STA 6};
                \node[sta_freq, below = of DATA_STA1, yshift = .2cm, font=\Large
                    ](DATA_STA2){DATA STA 1};
                \node[sta_freq, below = of DATA_STA2, yshift = .2cm, font=\Large
                    ](DATA_STA3){DATA STA 3};
                \node[sta_freq, below = of DATA_STA3, yshift = .2cm, font=\Large
                    ](DATA_STA4){DATA STA 5};
                \node[ap_freq, right = of DATA_STA1.east, minimum height = 2.5cm,
                    yshift = -.75cm, font=\Large](BlockACK2)
                    {Multi-\\STA\\Block\\Ack \\(BA)};
                \path[line](-2,-2.5) -- node[pos=.5, rotate=90,
                    anchor=south, font=\Large]{frequency}++(0,3.1);
                \path[line](-2,-2.5) -- node[pos=0.95,
                    anchor=north, font=\Large]{time}++(13.6,0);
                \path[sifs](BSRP_TF3.north east)--node[pos=0.5,
                    ]{SIFS}(BSR_STA3.north west);
                \path[sifs](BSR_STA3.north east)--node[pos=0.5, ]{SIFS}(BlockACK1.west);
                \path[sifs](BlockACK1.east)--node[pos=0.5,
                    ]{SIFS}(BASIC_TF3.north west);
                \path[sifs](BASIC_TF3.north east)--node[pos=0.5,
                    ]{SIFS}(DATA_STA3.north west);
                \path[sifs](DATA_STA3.north east)--node[pos=0.5, ]{SIFS}(BlockACK2.west);
                \node at (-1, -5) (first_table) [font=\Large] {
                        \begin{tabular}{|c|c|}
                            \hline
                            STA & OBO \\
                            \hline
                            1 & 15 \\
                            \hline
                            2 & 1  \\
                            \hline
                            3 & 2 \\
                            \hline
                            4 & 5 \\
                            \hline
                            6  & 7 \\
                            \hline
                            8 & 3\\
                            \hline
                        \end{tabular}
                    };
                \node at (first_table.north) [above, font=\Large ] {Before BSRP TF};
                \node [right = of first_table.east, font=\Large] (first_table) {
                        \begin{tabular}{|c|c|}
                            \hline
                            STA & OBO \\
                            \hline
                            1 & 12 \\
                            \hline
                            2 & 0  \\
                            \hline
                            3 & 0 \\
                            \hline
                            4 & 2 \\
                            \hline
                            6 & 4 \\
                            \hline
                            8 & 0\\
                            \hline
                        \end{tabular}
                    };
                \node at (first_table.north) [above, font=\Large ] {After BSRP TF};
                \draw[red, thick] (-2.1cm, -4.3cm) rectangle (3.2cm,-4.8cm);
                \draw[red, thick] (-2.1cm, -5.28cm) rectangle (3.2cm,-4.8cm);
                \draw[red, thick] (-2.1cm, -6.62cm) rectangle (3.2cm,-6.2cm);
            \end{tikzpicture}
        }
        \caption{Frame exchange sequence for OFDMA channel access with UORA. Tables
    in the bottom show the values of OFDMA back-off (OBO) for all STAs before
and after BSRP TF is received. STAs whose OBO reaches 0 are framed with red.}
    \label{fig:uora_ofdma}
\end{figure}

In this section, we describe the operation of UORA in detail. After the AP
successfully gains channel access through contention, it initiates UL
transmission by sending a Trigger Frame (TF). In this TF, the AP can designate
each RU for either RA, allowing all STAs to use it, or SA, restricting its use
to a single designated STA. While several types of TFs are defined in the
standard, we focus on two that are particularly relevant for describing the
operation of UORA: Buffer Status Report Poll (BSRP) TF and Basic TF. The BSRP TF
directs STAs to report their buffer status, while the Basic TF signals for UL
data transmission.
During the association stage, the AP sends the UORA parameter set, which
includes the OFDMA Contention Window ($OCW$) range defined by $EOCW_{min}$ and $EOCW_{max}$.
These parameters are transmitted in the management frames, and their values
can be adjusted as needed. After receiving these parameters, a STA uses them to
set the value of its $OCW$. When a STA receives a TF in which it is not
explicitly allocated any of the SA RUs, and the TF indicates that RA is
permitted, the STA may partake in UORA (e.g., when it needs to transmit a new
buffer status report). In such cases, it initializes
its $OCW$ to $OCW_{MIN} = 2^{EOCW_{min}} - 1$ and randomly selects an
initial OFDMA back-off ($OBO$) value within the range $[0, OCW]$. The maximum
$OCW$ value is determined as $OCW_{MAX} = 2^{EOCW_{max}} - 1$. Then upon receiving a new TF, it decreases its
$OBO$ counter by the number of RA RUs indicated in this TF. If the $OBO$ counter is less than or equal
to the number of RA RUs, the STA randomly selects one of advertised RA RUs
in the TF and uses it to transmit. If the transmission is successful, the STA resets its
OFDMA contention window (OCW) to $OCW_{MIN}$. In the event of a failed
transmission, the OCW is doubled each time until it reaches an upper bound of
$OCW_{MAX}$. 

Figure \ref{fig:uora_ofdma} illustrates the UORA frame exchange sequence. The AP
initiates an UL OFDMA transmission by sending a BSRP TF, which includes three RA
RUs (denoted by AID 0) and a SA RU assigned to STA 5. This BSRP TF prompts STAs
to report their buffer status, enabling the AP to identify which STAs require
resources. After one Short Interframe Space (SIFS), a STA transmits a BSR using
either a randomly selected RA RU or an SA RU assigned to it. A STA is eligible
to select an RA RU if its $OBO$ value is less than or equal to the number of RA
RUs advertised in the BSRP TF. STAs 2, 3, and 8 meet this criterion.
Consequently, after a SIFS, they randomly select and transmit using one of the
RA RUs. Specifically, STA 3 selects the third RA RU and successfully transmits,
while STAs 2 and 8 transmit using the same RA RU, resulting in a collision
(shown as the shaded area), leaving one unused RU. Following this, the AP
acknowledges the transmission by sending a Multi-STA Block ACK after a SIFS. STA
3, which transmitted successfully, randomly selects a new OBO value from the
range $[0, OCW = OCW_{MIN}]$, whereas STAs with unsuccessful transmissions
choose their new OBO value from $[0, 2^{OCW} - 1]$, up to a maximum of
$OCW_{MAX}$. After another SIFS, the AP allocates RUs to STAs that have reported
having data to transmit (STAs 3 and 5). The AP can also allocate resources to
STAs that have previously reported non-zero buffer statuses, or to STAs that it
thinks might have data for transmission but failed to deliver their buffer
statuses. For example, on the picture the AP also allocates resources to STAs 1
and 6. It is also allowed to assign some of the RUs for random access but the
overhead due to collisions of the data packets is usually significantly higher
than that for BSRs. Note that RUs can be of different sizes depending on the
needs of the STAs, but for simplicity, they are considered the same. This
allocation is communicated through the Basic TF. In the BSRP Trigger Frame (TF),
the RU allocation identifies the RUs assigned for STAs to transmit Buffer Status
Reports. In contrast, the Basic TF specifies RUs for STAs to transmit UL data
packets. Following one more SIFS, these STAs transmit on their respective RUs.
To ensure synchronized transmission, smaller payloads are padded to match the
size of the largest payload. Finally, the AP sends a Multi-STA Block ACK after a
SIFS, thereby concluding the UL OFDMA transmission with RA RUs.

To mitigate contention and improve airtime fairness in densely populated
networks, the Wi-Fi 6 standard introduced the MU EDCA Parameter Set, which
includes EDCA parameters such as contention window and Arbitration Inter-Frame
Space Number (AIFSN) that can be used by STAs after participating in uplink (UL)
OFDMA transmission. This way, the network can be configured so that the AP
exerts greater control over UL transmissions, while the STAs themselves compete
less aggressively, or do not compete at all, for the channel. The AP announces
the MU-EDCA Parameter Set through management frames. When this set includes an
AIFSN value of zero, it signals the STAs to disable EDCA-based contention
entirely. In such cases, the AP fully orchestrates UL transmissions, and STAs do
not contend for medium access. In addition to EDCA parameters, the MU EDCA
Parameter Set includes a validity timer, known as the MU EDCA timer. This timer
dictates how long a STA should apply the received parameters. The timer is reset
each time the STA successfully transmits data via OFDMA and receives a
corresponding Block Ack from the AP. If the timer expires without a successful
transmission, the STA reverts to its default EDCA settings. As long as the STA
continues to be scheduled by the AP, it retains the updated parameters.

Many studies proposed various modifications to UORA in order to improve its
performance. In particular, modifications have been proposed for the selection
of the $OBO$
counter~\cite{rehman2023collision, kim2021ofdma}, back-off countdown
procedure~\cite{kosek2022efficient, kosek2022improving}, and collision
resolution~\cite{avdotin2019enabling}. Xie \textit{et al.}~\cite{xie2020multi}
proposed to reduce collisions among stations contending for the same RU by
applying busy tone arbitration. Multiple papers studied combining UORA with
other mechanisms that emerged in new Wi-Fi standards, such as multi-link
operation~\cite{jin2024enhancing}, and target wake
time~\cite{chen2020scheduling}. Furthermore, some
papers~\cite{bhattarai2019uplink, avdotin2020resource} proposed scheduling
algorithms that leverage both random and scheduled access procedures.

Despite the considerable number of papers listed studying UORA, none of these
papers provide a validated implementation of UORA in a trustworthy network
simulation platform. To our knowledge, the only publicly available
validated implementation has been developed by Naik et al. in~\cite{1_Naik}.
However, in the implementation of Naik \textit{et al.}, resource allocation
occurs only at the beginning of each UL OFDMA transmission when polling STAs for
their buffer status. As a result, collided or idle RUs during polling are not
utilized for data transmission, leading to a decrease in the expected
performance of UORA. Additionally, signaling of the UORA parameter set to STAs
via management frames has not been implemented. The signaling of
parameters enhances flexibility and control over parameter configuration and
adjustments. For example, in our implementation, $OCW_{min}$ and $OCW_{max}$ can
be dynamically modified based on the traffic density of the deployment.
Consequently, our work does not incorporate any component of the aforementioned
implementation.

\tikzstyle{rec_block} = [rectangle, draw, thick, text centered, minimum height=1cm,
minimum width = 1.5cm, outer sep = 0pt]
\tikzstyle{rec_block_ocwrange} = [rectangle, draw, text centered, minimum
height=.5cm,
minimum width = .5cm, outer sep = 0pt]
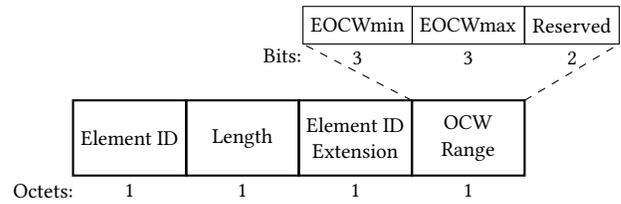
\begin{figure}[tb]
    \Description{Empty}
    \centering
    \begin{tikzpicture}[font=\small, on grid, auto, label position=center, align
        = flush center]

        \node[rec_block](ElementID) {Element ID};
        \node[rec_block, right = of ElementID, xshift=.5cm](Length) {Length};
        \node[rec_block, right = of Length, xshift=.5cm](ElementIDExt) {Element
            ID\\Extension};
        \node[rec_block, right = of ElementIDExt, xshift=.5cm](OCWRange) {OCW\\Range};

        \node[below=of ElementID, yshift=.3cm](1){1};
        \node[below=of Length, yshift=.3cm](2){1};
        \node[below=of ElementIDExt, yshift=.3cm](3){1};
        \node[below=of OCWRange, yshift=.3cm](4){1};
        \node[below=of ElementID, yshift=.3cm, xshift=-1.15cm](octets){Octets:};

        \node[rec_block_ocwrange, above = of OCWRange, yshift=.5cm](EOCWmax) {EOCWmax};
        \node[rec_block_ocwrange, left = of EOCWmax, xshift=-.47cm](EOCWmin) {EOCWmin};
        \node[rec_block_ocwrange, right = of EOCWmax, xshift=.38cm](Reserved)
            {Reserved};

        \node[below=of EOCWmin, yshift=.56cm](OC1){3};
        \node[below=of EOCWmax, yshift=.56cm](OC2){3};
        \node[below=of Reserved, yshift=.56cm](OC3){2};
        \node[below=of EOCWmin, yshift=.6cm, xshift=-1cm](Bits){Bits:};

        \draw[dashed] (OCWRange.north east) -- (Reserved.south east);
        \draw[dashed] (OCWRange.north west) -- (EOCWmin.south west);

    \end{tikzpicture}
    \caption{UORA Parameter Set element format}
    \label{fig:param_set}
\end{figure}
    
\section{UORA Implementation in NS-3}
\label{sec:uora_impl}

In this section, we provide a high-level overview of the classes of the Wi-Fi
module that were created or modified to implement
the UORA functionality, explaining their interactions in enabling this feature
in both the AP and STAs. We conclude by highlighting the aspect of UORA usage
that is yet to be supported.

Our UORA implementation extends the ns-3 IEEE~802.11ax Wi-Fi models by adding new
classes and modifying existing ones. Figure \ref{fig:ns3_uoraImp} depicts an
abridged view of the ns-3 IEEE~802.11ax Wi-Fi models where components with a grey
background indicate the modified and newly added classes. We introduce a
new class, \emph{UoraParameterSet}, that implements the UORA Parameter Set
element. The UORA Parameter Set element consists of an octet for each of the
element id, the length, the extension id and the $OCW$ range, as illustrated in
Figure~\ref{fig:param_set}. The first three bits of the most significant octet,
$OCW$ range, is $EOCWmin$ and the next three bits represent $EOCWmax$, with the
final two bits reserved for future use. Additionally, we modified the
\emph{MgtHeaders} class to incorporate the UORA parameters in the management
frames. The \emph{HeConfiguration} class configures the $OCWmin$ and the
$OCWmax$ values on the AP side. The configured values are subsequently broadcast
to the STAs through management frames.

The \emph{ApWifiMac} and \emph{StaWifiMac} classes manage beacon generation,
probing, and association functionality in Wi-Fi. Consequently, these classes
have been modified to incorporate UORA parameters into the management frames.
Furthermore, the process of updating a STA's OCW value after a successful
transmission on a randomly selected RU or a collision caused by multiple STAs
transmitting on the same RU, along with selecting a new $OBO$ value in either
case, is implemented in \emph{QosTxop}. More specifically, the
\emph{QosTxop::UpdateObo} function resets the $OCW$ value to $OCW_{min}$ and
assigns a new $OBO$
value following a successful transmission. In the case of a failed transmission
attempt, the \emph{QosTxop::UpdateFailedOcw} function updates both the $OCW$ and
$OBO$
values accordingly. Additionally, we have extended the
\emph{QosTxop::StartMuEdcaTimerNow} function to support disabling EDCA for the entire
simulation duration, allowing for precise modeling of contention-free STA
behavior. The \emph{HeFrameExchangeManager} is
responsible for managing the frame exchange sequence of UORA, as illustrated in
Figure \ref{fig:uora_ofdma}, ensuring that each step in the sequence is executed
at the appropriate time. For instance, it enforces that the AP sends a Multi-STA
Block ACK after one SIFS when it receives a BSR from a STA on an RA RU. 
Additionally, it interacts with the \emph{RrMultiUserScheduler} object to
allocate RUs to STAs during scheduled access while reserving some RUs for RA.
The \emph{RrMultiUserScheduler} class has been extended to prioritize the
rescheduling of unused RUs from the BSRP/BSR exchange, whether due to collisions
or lack of selection by STAs, for packet transmission in the Basic TF, thus
improving resource efficiency. However, it also accommodates scenarios where
unused RUs from BSR transmission remain unscheduled in the Basic TF. In UORA,
the same RU may be "shared" by multiple STAs even in the absence
of MU-MIMO support. This occurs when multiple STAs randomly select the same RU
for uplink transmission. In such scenarios, the \emph{HePhy} class has been
extended to synchronize to the first successfully detected PPDU, while all other
concurrently transmitted PDUs within the same RU are treated as interference.

Unassociated STAs can send association requests in RA RUs, provided they receive
a TF with the AID 2045 of an RA RU. However, this feature of UORA is not yet
implemented. The code for this work is available at \cite{UORAns3}.

\tikzstyle{block} = [rectangle, draw, text centered, minimum height=.5cm,
minimum width = 1cm, outer sep = 0pt]
\tikzstyle{arrow} = [draw, -{latex}]
\tikzstyle{dotted-arrow} = [draw, -{Stealth[open, width=3mm]}, dashed]
\tikzstyle{dotted-arrowreg} = [draw, -{>[open]}, dashed]

\tikzstyle{diamond-arrow} = [draw, -{Diamond[open,scale=1.5]}]
\tikzstyle{dottedbox} = [draw, dashed, thin]
\begin{figure}[tb]
    \Description{Empty}
    \centering
    \scalebox{0.8}{
        \begin{tikzpicture}[node distance = 1.3cm, font=\small, on grid, auto,
            label position=center, align = flush center]
            \node [block, fill=gray!50] (high_mac) {ApWifiMac/StaWifiMac};
            \node [block, below=of high_mac.west, anchor=west, fill=gray!50]
                (qosTxop){QosTxop};
            \node [block, right=1.7cm of qosTxop] (MacRxMiddle) {MacRxMiddle};
            \node [block, below= of $(qosTxop)!0.5!(MacRxMiddle)$, fill=gray!50,
                yshift = -.3cm](HeFrameXchange){HeFrameExchangeManager};
            \node [block, below =of HeFrameXchange, minimum width=2cm,
                fill=gray!50, ] (HePhy){HePhy};
            \node [block, right= of high_mac.east, fill=gray!50] (HeConf)
                {HeConfiguration};
            \node [block, below = of HeConf, yshift=.3cm, fill=gray!50]
                (MgtHdrs){MgtHeaders};
            \node [block, below = of MgtHdrs, yshift=.3cm, fill=gray!50]
                (UoraPara){UoraParameterSet};
            \node [block, left = of qosTxop.west, xshift=-.6cm,
                ](ChannelAccessMgr) {ChannelAccess\\Manager};
            \node [block, right = of HePhy.west, xshift=2cm,
                ](PhyEntity) {PhyEntity};
            \node [block, right = of
                HeFrameXchange.east,xshift=.3cm,
                fill=gray!50](rrShed){RrMultiUser\\Scheduler};
            \path [arrow] ([xshift=-.85cm]high_mac.south) --node[pos=0.3,
                left=-2pt]{Queue (packet)} (qosTxop.north);
            \path [arrow] (MacRxMiddle.north) --node[pos=0.5,
                right=-2pt]{Receive} ([xshift=0.84cm]high_mac.south);
            \path [arrow] (qosTxop.south) --node[pos=0.7,
                left=-2pt]{StartTransmission} ([xshift =
                -0.85cm]HeFrameXchange.north);
            \path [arrow] ([xshift = 0.85cm]HeFrameXchange.north)
                --node[pos=0.5, right=-2pt]{Receive} (MacRxMiddle.south);
            \path [arrow] ([xshift = -0.7cm]HeFrameXchange.south)
                --node[pos=0.5, left=-2pt]{Send} ([xshift =
                -0.7cm]HePhy.north);
            \path [arrow] ([xshift = 0.7cm]HePhy.north) --node[pos=0.6,
                right=-2pt]{ReceiveOk/ReceiveError} ([xshift =
                0.7cm]HeFrameXchange.south);
            \path [dotted-arrowreg] (high_mac.east)-- node[pos=0.5,
                above]{<<use>>} (HeConf.west);
            \draw [->, dashed] (high_mac.east) --++ (.8, 0) |-++ (0, -1) --
                 (MgtHdrs.west);
            \draw [{latex[width=0.5cm]}-{latex[width=0.5cm]}]
                (HeFrameXchange.east) --node[above = 5pt]{SelectTxFormat}
                (rrShed.west);
            \path [dotted-arrowreg] (MgtHdrs.south) --
                node[]{<<use>>}(UoraPara);
            \path [diamond-arrow] (HePhy.east) -- (PhyEntity.west);
            \path [arrow] (ChannelAccessMgr.east) --node[pos=0.5,
                below=-11pt]{Notify\\AccessGranted} (qosTxop.west);
            \draw [-, dashed] (HeFrameXchange.west) -- node[pos=0.5,
                below=]{Listener} (-4.45, -3.15);
            \path [dotted-arrow] (HePhy.west) -| node[pos=0.3,
                below=]{Listener} 
                (ChannelAccessMgr.south);
            \node[fit=(qosTxop)(MacRxMiddle), dottedbox] {};
        \end{tikzpicture}
    }
    \caption{Abridged ns-3 802.11ax Wi-Fi models: components with a grey
    background represent modified or newly introduced classes in our
    Implementation}
    \label{fig:ns3_uoraImp}
\end{figure}
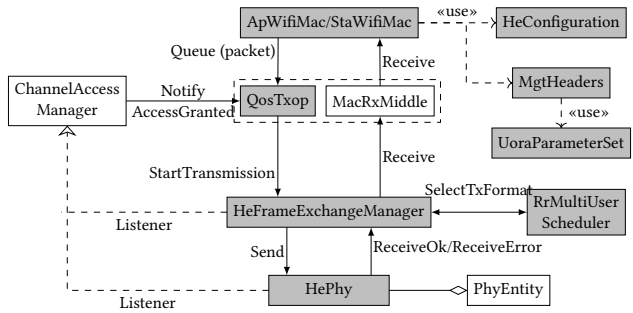

\section{Validation of Implementation}
\label{sec:validation}

To validate our UORA implementation, we run simulations in ns-3
and collect relevant performance metrics. The simulation
results are then compared against the analytical model proposed by Naik et al.
\cite{1_Naik}. This comparison helps verify that our UORA model aligns with
theoretical expectations.

\subsection{Simulation Setup}
In our simulation, we model a single Basic Service Set (BSS) consisting of
multiple STAs and a single AP, both compliant with the IEEE~802.11ax standard.
The STAs generate a constant bit rate (CBR) flow of UDP packets with the AP as
destination. Downlink traffic generation is deliberately disabled, as this work
focuses on UORA, which is suitable for only UL transmissions. Packet size is
configured in such a way that the transmission always fits within the Transmit
Opportunity (TXOP) allocated for data transmission over a 26-tone RU, without
introducing additional gaps that could negatively impact the expected throughput.

Furthermore, to maintain a continuous frame queue, the inter-packet generation
interval is set to one-forth of the TXOP duration, ensuring that packets are
consistently available for transmission thus maximizing channel efficiency. The
resulting UDP data rate is approximately 26.58 Mbps per uplink STA. All traffic
is assigned to the Voice (VO) Access Category (AC), meaning that only a single
queue is active on each device. The Multi User EDCA (MU-EDCA) timer parameter
for this queue is configured to the simulation duration, with the AIFSN set to
zero---effectively disabling EDCA for the duration of the experiment. This
prevents STAs from transmitting packets using EDCA when the AP schedules
resources with UL OFDMA. Additionally, the Beacon interval is set to a
relatively higher value of \qty{204800}{\micro\second} (vs. default of 100 TUs)
to minimize the impact of Beacon frames on the throughput measurement, as the
analytical model \cite{1_Naik} does not consider the occasional Beacon
transmissions. PHY-level aggregation is enabled, which is essential for
achieving the expected throughput when utilizing larger RUs, such as 106-tone
allocations, by effectively maximizing the use of available transmission
opportunities.

We ensure that the probability of packet loss due to channel errors is
negligible by transmitting at sufficiently high power, which guarantees that all
STAs are within the AP's range. As a result, packet loss only occurs when two
or more STAs simultaneously transmit in the same randomly chosen RU.
Furthermore, we use the smallest RU type (i.e., 26-tones) under 20 MHz bandwidth
and 106-tone RUs when utilizing 80 MHz bandwidth to ensure an equitable
distribution of resource among associated STAs and maximize the number of
simultaneous transmissions. To ensure statistical robustness, each experiment
was repeated five times.
Table~\ref{table:sim_para} lists the parameter values used in the experiments.

\begin{table}
    \caption{List of simulation parameters}
    \label{table:sim_para}
    \centering
    \renewcommand{\arraystretch}{1.3} 
    \begin{tabular}{p{5cm} c}
        \toprule
        \textbf{Parameter} & \textbf{Value} \\
        \midrule
        Carrier frequency & \qty{5}{\giga\hertz} \\
        Bandwidth & \{20, 80\}~\si{\mega\hertz} \\
        Guard Interval & \qty{0.8}{\micro\second} \\
        MCS Index & 8 \vspace{.2cm}\\
        Resource Unit Type & \pbox{2.5cm}{26-tone~in~\qty{20}{\mega\hertz}\\
        106-tone~in~\qty{80}{\mega\hertz}} \vspace{.2cm}\\
        Transmit Opportunity & \qty{2080}{\micro\second} \\
        Beacon Interval & \qty{204800}{\micro\second} \\
        OCW$_{min}$ & 31 \\
        OCW$_{max}$ & 127 \\
        AP Access Request Interval & \qty{124}{\micro\second} \\
        EDCA Access Category & VO \\
        Uplink Payload Size & \qty{1700}{\byte} \\
        Traffic profile & Full buffer \\
        Duration of Simulation & \qty{15}{\second} \\
        Number of Simulation Runs & 5 \vspace{.2cm} \\
        Number of Contending STAs & \pbox{2.4cm}{\{ 9, 18, 27, 36, 45, 54, 63,
                72, 81, 90, 99 \}} \\
        \bottomrule
    \end{tabular}
\end{table}

\subsection{Discussion of Results}

\begin{figure}
    \centering
    \begin{subfigure}[b]{0.81\linewidth}
    \includegraphics[width=\linewidth]{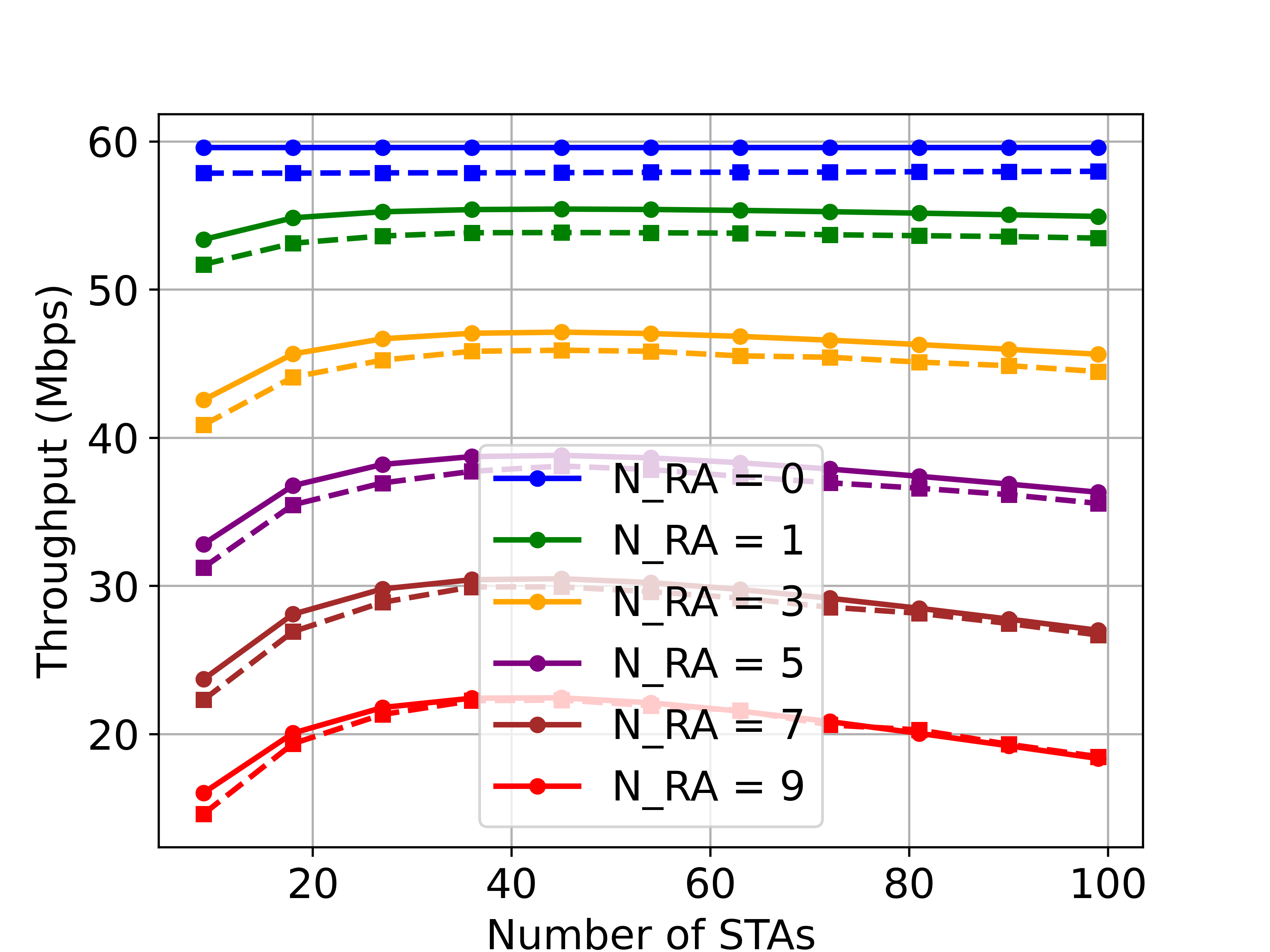}
    \caption{20 MHz Bandwidth}
    \label{fig:result_20Mhz}
  \end{subfigure}
  \begin{subfigure}[b]{0.81\linewidth}
    \includegraphics[width=\linewidth]{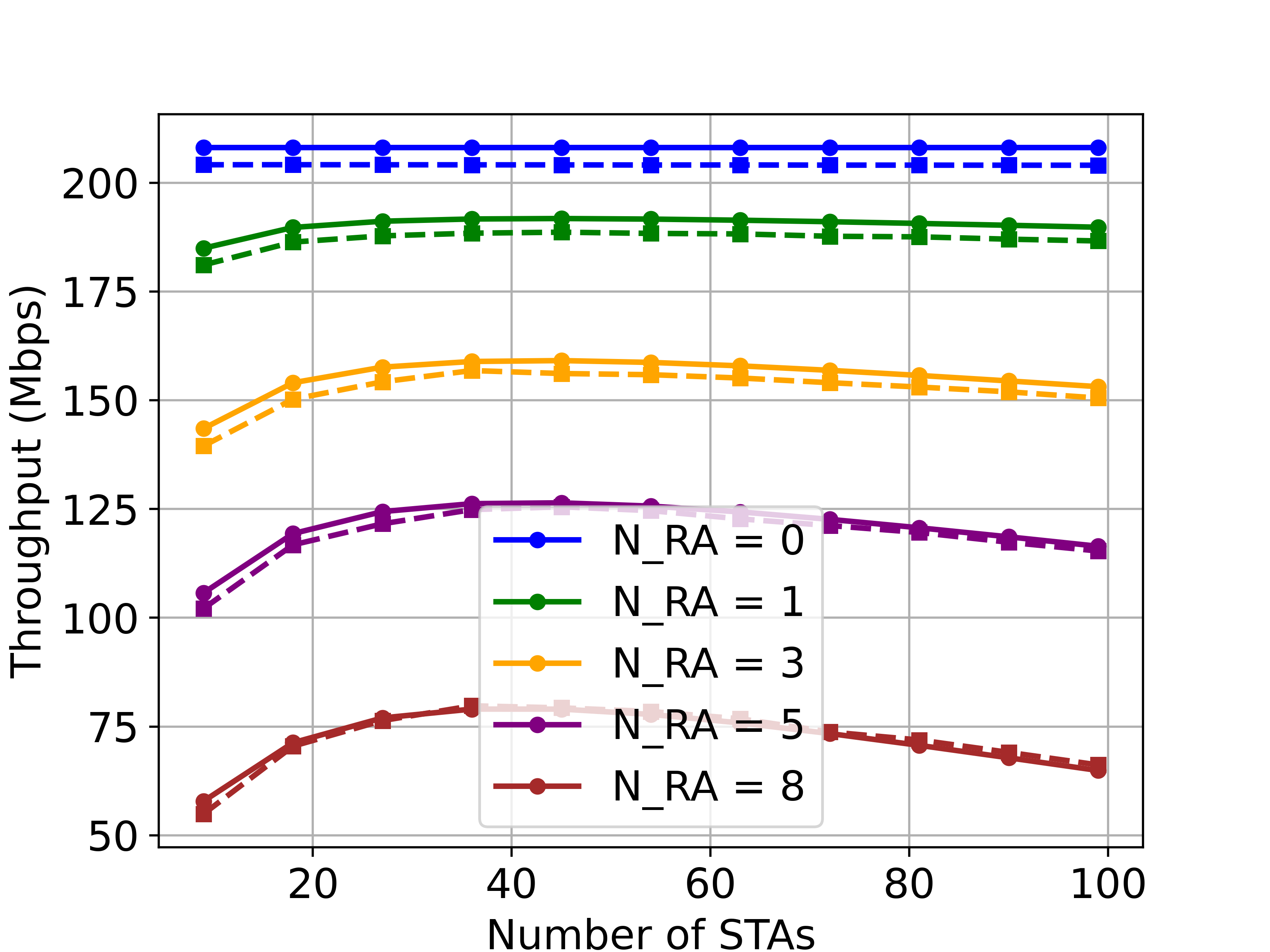}
    \caption{80 MHz Bandwidth}
    \label{fig:result_80Mhz}
  \end{subfigure}
    \Description{Solid lines depict the results from the analytical model and
    the dashed lines represent the simulation results.}
    \caption{Validation of UORA implementation: throughput as a function of the
    number of STAs for different number of RA RUs. Solid lines depict the
results from the analytical model~\cite{1_Naik} and the dashed lines represent
the simulation results.}
    \label{fig:mmodel_sim_throughput}
\end{figure}

In this section, we compare the simulation results of our UORA implementation
with those obtained from the analytical model proposed by Naik \textit{et al.}
\cite{1_Naik}, using the same parameters. The evaluation focuses on throughput
as the key performance metric. Figure~\ref{fig:mmodel_sim_throughput} depicts
application-level throughput for different numbers of STAs, considering varying
number of RA RUs (i.e., $N_{RA}$). Figure~\ref{fig:result_20Mhz} presents the
results for 20~MHz bandwidth, while Figure~\ref{fig:result_80Mhz} shows the
corresponding results for 80~MHz bandwidth. As expected, the performance
improvement is approximately fourfold. The solid lines represent the analytical
results, while the dashed lines correspond to the simulation results.

The simulation throughput, while closely following the trend of the mathematical
model, is slightly lower, particularly in scenarios with few RA RUs. This
discrepancy may be attributed to the fact that, although the AP does not have downlink traffic in
the considered scenario, it must still obtain channel access before it can
initiate an uplink transmission by sending a Trigger Frame. Importantly, the AP
does not compete with STAs for channel access, as EDCA is disabled on the STAs.
However, in the ns-3 implementation of OFDMA, the AP can only initiate a new
channel access—i.e., transmit a Trigger Frame—after a specific period, defined
by the attribute ns3::MultiUserScheduler::AccessReqInterval, has elapsed since
the last successful channel access. As a result, there are pauses between the
end of one UL OFDMA exchange and the start of the next. These idle periods
introduce additional delays that are not modeled in the analytical framework and
contribute to the discrepancy between the simulation results and the
mathematical model. Moreover, the inclusion of an additional TF (i.e., Basic TF)
in the UL OFDMA transmission further reduces the achievable throughput.

\section{Conclusion}
\label{sec:conclusion}

In this paper, we presented a standard-compliant implementation of UORA
that is compatible with the ns-3 version 3.38 simulator. By
incorporating a standalone resource scheduler, our implementation provides a
more accurate and flexible framework for evaluating UORA’s performance in dense
networks, where efficient UL scheduling is critical. We validate our
implementation by comparing its results to those of an analytical model from
literature. In our future research, we plan to design the mathematical model of
the scheduler that decouples resource allocation for BSRs and data
transmissions. We will use this implementation to validate this model.
To further explore the discrepancy between the mathematical and simulation
results, we will, for instance, vary the
ns3::MultiUserScheduler::AccessReqInterval parameter, align EDCA configurations
as closely as possible with the analytical model, and trace corresponding events
during simulation.

\section*{ACKNOWLEDGEMENTS}
This research was partially funded by the ICON project VELOCe (VErifiable,
LOw-latency audio Communication), realized in collaboration with imec, with
project support from VLAIO (Flanders Innovation and Entrepreneurship). Project
partners are imec, E-Bo Enterprises, Televic Conference, and Qorvo. The research
was supported by the FWO WaveVR project (G034322N).

\bibliographystyle{ACM-Reference-Format}
\bibliography{uoraImpl}

\end{document}